\documentclass[%
 reprint,
 amsmath,amssymb,
 aps,
nofootinbib
]{revtex4-2}

\usepackage{graphicx}
\usepackage{dcolumn}
\usepackage{bm}
\usepackage{hyperref}
\usepackage[compat=1.1.0]{tikz-feynman} 
\usepackage{physics}

\newcommand{\kb}{$k_\textrm{B}$}
\newcommand{\mEr}{m_\textrm{Er}}
\newcommand{\mLi}{m_\textrm{Li}}
\usepackage{soul}
\usepackage{braket}
\usepackage{upgreek}

\begin{document}

\preprint{APS/123-QED}

\title{Creation of a degenerate Bose-Bose mixture of erbium and lithium atoms}

\author{Jasmine Kalia}
\thanks{These two authors contributed equally, and ordering is alphabetical.}
\author{Jared Rivera}
\thanks{These two authors contributed equally, and ordering is alphabetical.}
\author{Rubaiya R. Emran}
\author{William J. Solorio Hernandez}
\author{Kiryang Kwon}
\author{Richard J. Fletcher}
\email{rfletch@mit.edu}

\affiliation{%
 MIT-Harvard Center for Ultracold Atoms, Research Laboratory of Electronics, and Department of Physics, Massachusetts Institute of Technology, Cambridge, Massachusetts 02139, USA 
}%

\begin{abstract}
We report the realization of a degenerate mixture of $^{166}$Er and $^{7}$Li atoms in their energetically lowest spin states. 
The two species are sequentially laser-cooled and loaded into an optical dipole trap, then transported to a glass cell and simultaneously evaporated to degeneracy. 
Er serves as the coolant for Li, and we observe efficient sympathetic cooling facilitated by a large interspecies elastic scattering cross section. 
Three-body losses are found to be small, making this platform promising for the study of interacting mixtures with large mass imbalance.
\end{abstract}

\date{\today}

\maketitle


\section{\label{sec:level1}Introduction}
Ultracold atomic mixtures of different elements provide qualitatively new tools for engineering and exploring many-body physics~\cite{Baroni:2024mixRev}. 
On one hand, they enable the study of heterogeneous system phenomena including impurity and few-body physics~\cite{Yan:2020bosePolaron,Massignan:2014polaronReview,Nishida:2016kondo,Naidon:2017efimovReview}, 
quantum fluid mixtures~\cite{Tuoriniemi:2002heliumMixture,FerrierBarbut:2014mixedSuperfluids},
and imbalanced fermionic pairing relevant to nuclear science~\cite{Barrois:1977quarkSC}, condensed matter~\cite{Casalbuoni:2004inhomoReview}, and astrophysics~\cite{Sedrakian:1997imbalAstro}. 
On the other, their additional richness provides new possibilities such as mixed dimensional confinement via species-selective potentials~\cite{Nishida:2009efimovConfinement,Lamporesi:2010mixedDim} and emergent mediated interactions between particles~\cite{DeSalvo:2019fermMed,Baroni:2023intPolarons}. They can also be used to construct the building blocks of more complex systems such as dipolar molecules~\cite{Carr:2009molRev,Langen:2024moleculeReview,Cornish:2024molRev}, with applications to many-body physics~\cite{Baranov:2012dipolarReview}, quantum chemistry~\cite{Karman:2024moleculesChemistry}, and quantum computation~\cite{demille:2002qcMolecules,Karra:2016qcMagMol}.

Recently, alkali-lanthanide mixtures have emerged as a promising new frontier~\cite{ravensbergen2018production,schafer2022feshbach,schaffer2023fermionicEr,xie2025feshbach}, offering several advantages. 
First, they feature large mass ratios, which are favorable for directions including controlling the energy and length scalings of Efimov states~\cite{Pires:2014efimovMassImb,Tung:2014efimovMassScaling,Naidon:2017efimovReview} and a substantially enhanced critical temperature for exotic superfluidity in Bose-Fermi mixtures~\cite{Wu:2016topological,Caracanhas:2017}. 
Second, the availability of numerous stable isotopes allows access to all combinations of quantum statistics. 
Third, they have been predicted theoretically~\cite{gonzalez2015erliFeshbach} and found experimentally~\cite{schafer2022feshbach,schaffer2023fermionicEr,xie2025feshbach} to exhibit convenient Feshbach resonance~\cite{Chin:2010FeshbachReview} spectra that are sufficiently sparse to avoid overly deleterious losses yet frequent and broad enough for control of interspecies interactions. 
In addition, the interplay of long-range magnetic dipolar forces with short-range contact interactions has been predicted to underlie spontaneous pattern formation~\cite{bland:2022mixtureSupersolids} and to stabilize supersolid phases at the mean-field level~\cite{Li:2022mixSupersolid}, while magneto-association via Feshbach resonances promises heteronuclear molecules featuring both electric and magnetic dipole moments~\cite{Karra2016magMoleculesTheory,Finelli:2024licrMolecules}. 

To date, the only degenerate candidate realized is a Fermi-Fermi mixture of Dy-K~\cite{ravensbergen2018production}, while in the past few years thermal mixtures of Er-Li~\cite{schafer2022feshbach,schaffer2023fermionicEr} and Dy-Li~\cite{xie2025feshbach} have been reported. 
A closely-related combination, Li-Cr, was recently brought to degeneracy~\cite{Ciamei:2022licrDegen,Finelli:2024licrMolecules}, and Dy-Er mixtures have also been achieved~\cite{Trautmann:2018dyermixture}. Li-Yb mixtures have been successfully cooled~\cite{hansen:2011degenerateLiYb,Hara:2011ybliDegenerate} and provide a comparable mass ratio, but control of interactions is challenging~\cite{Dowd:2015liybMagneticField,Schafer:2017liybMagnetic,green:2020feshbach}.

\begin{figure*}[t]
\includegraphics[width=\textwidth]{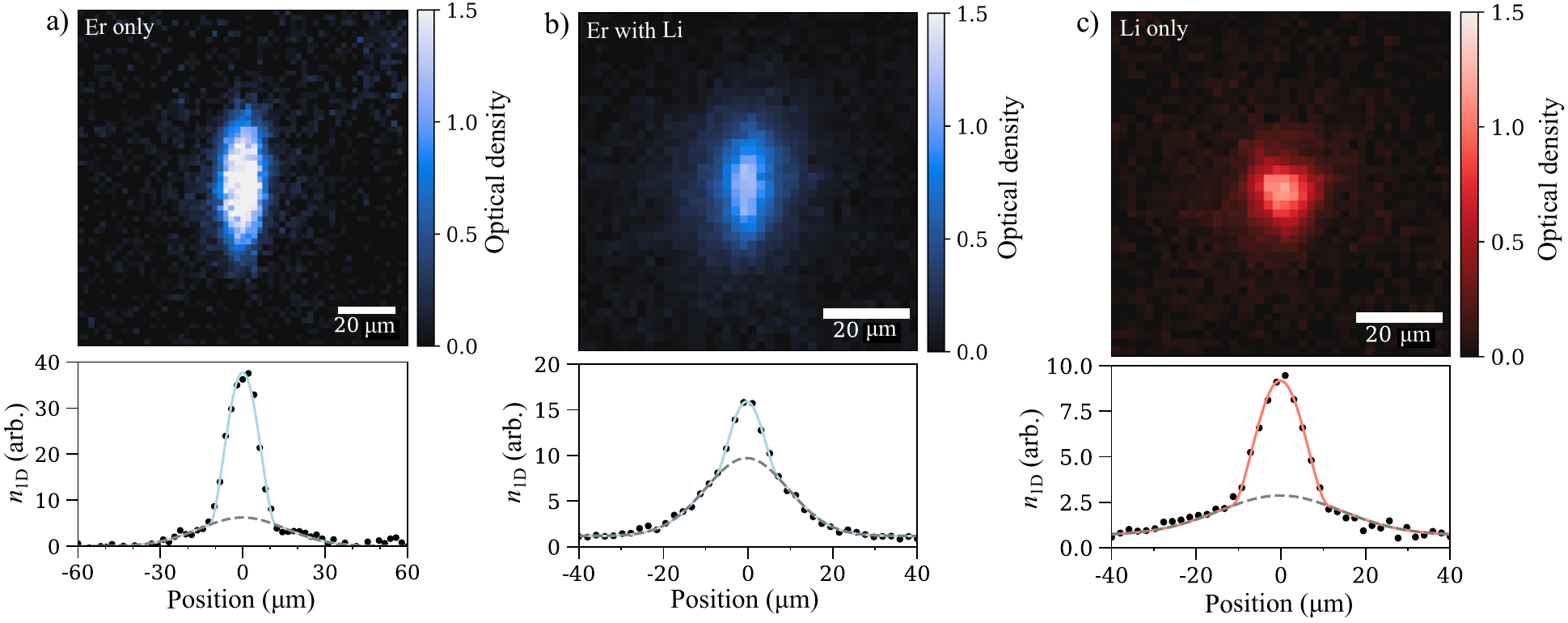}
\caption{Bose-Einstein condensation of $^{166}$Er and $^7$Li. Er is cooled via forced evaporation in an optical trap, and sympathetically cools the Li cloud. The top panels show absorption images of condensed clouds and the bottom panels show the atomic density integrated along both the imaging and vertical axes, $n_{1\textrm{D}}$. The data are fit with a bimodal distribution capturing the thermal and condensed atoms, shown in dashed and solid lines respectively. (a) In the absence of Li, we produce condensates of $1.5\times 10^4$ Er atoms (time-of-flight $12~$ms). (b)-(c) Cooling an Er-Li mixture yields simultaneous condensates of both species, with $\sim 10^3$ atoms in each (time-of-flight $8~$ms and $0.8~$ms, respectively). }
\label{fig:dualBEC}
\end{figure*}

In this article, we describe the production of a degenerate mixture of bosonic $^{166}$Er and $^7$Li atoms, which is depicted in Fig.~\ref{fig:dualBEC}. After sequential laser cooling of each species, the mixture is loaded into an optical dipole trap, transported to a glass cell, and evaporatively cooled to degeneracy.
At the trapping wavelength of $1064~$nm, the polarizabilities of Er and Li are $\alpha_\textrm{Er}\approx 166~$a.u.~\cite{Becher:2018polarizability} and $\alpha_\textrm{Li}\approx 270~$a.u.~\cite{UDportal}, respectively. This means that Er experiences a shallower trap and therefore serves as a coolant for Li, and we observe highly efficient sympathetic cooling with minimal three-body losses. An interspecies thermalization measurement reveals a large elastic scattering length $a_{\textrm{ErLi}} = 100 \pm 10~a_0$, where $a_0$ is the Bohr radius, providing a promising route toward studying the interplay of strong interactions and large mass imbalance.

\section{\label{sec:level1}Optical cooling, trapping, and transport}
Here, we give an overview of the initial cooling and trapping stages of the experiment. 
Sec.~\ref{subsec:laserCooling} describes the laser cooling of each species and their collection in magneto-optical traps (MOTs), while Sec.~\ref{subsec:loadingTransportBeam} discusses their subsequent loading into an optical trap. Sec.~\ref{subsec:transport} outlines their transport to a glass cell, where they are evaporatively cooled to degeneracy. A more detailed overview of the apparatus itself is given in Appendix~\ref{appendix:apparatus}. 

\subsection{\label{subsec:laserCooling}Laser cooling}
The Er atomic beam is produced by a commercial effusion oven and transversely cooled using the broad $401~$nm transition. 
The majority of the Er atoms then transit a Li oven, from which an atomic beam of both species emerges.
The atoms then pass through a shared Zeeman slower, which employs a single magnetic field profile appropriate for both species.

Lithium atoms are collected in a MOT operating on the D$_2$ line at $671~$nm. 
The magnetic quadrupole field has a gradient of $20~$G$/$cm along the strong axis, and the cooling and repump beams each have a waist of $14~$mm. 
The cooling light is red-detuned by $5.9 ~\Gamma_{\textrm{D}_2}$ from the $\ket{F=2}\rightarrow \ket{F'=3}$ transition while the repump light is red-detuned by $11 ~\Gamma_{\textrm{D}_2}$ from the $\ket{F=1}\rightarrow \ket{F'=2}$ transition, and both  have a peak intensity of $\sim 0.5~I_{\textrm{s,D}_2}$. 
Here $\Gamma_{\textrm{D}_2}=2\pi\times 5.87~$MHz and $I_{\textrm{s,D}_2}=2.54~$mW$/$cm$^2$ are the natural linewidth and saturation intensity of the D$_2$ transition.
If the Li number is prioritized, we can collect $\sim 10^9$ atoms in the MOT. However, the subsequent sympathetic cooling of lithium by erbium is highly efficient and we find that the Er MOT loading rate benefits from a reduced Li atomic flux. We therefore operate the Li oven at a relatively cool temperature and typically load for several seconds at a loading rate of $~7\times10^6~$s$^{-1}$. 
After loading, we compress the Li MOT by ramping the detunings of the cooling and repump light to $2.4~ \Gamma_{\textrm{D}_2}$ and $8.4~ \Gamma_{\textrm{D}_2}$, respectively, over $25~$ms. 
This reduces the temperature of the trapped atoms from $1~$mK to $270~\upmu$K. 

Finally, we extinguish the MOT light and quadrupole field and perform gray molasses cooling on the D$_1$ line.
The gray molasses cooling beams each have a waist of $2.3~$mm, a central intensity of $ 13~I_{\textrm{s,D}_1}$, and are detuned by $2.4~  \Gamma_{\textrm{D}_2}$ from the $\ket{F=2}\rightarrow\ket{F'=2}$ transition. Here $I_{\textrm{s,D}_1}=7.59~$mW$/$cm$^2$ is the saturation intensity of the D$_1$ line. The light is oppositely circularly polarized to the MOT beams with which they co-propagate. After switching off the Li MOT, we wait $100~\upmu$s for transient magnetic fields to decay and then apply a cooling pulse of duration $2.7~$ms. We use a resonant EOM operating at $801.5~$MHz to add repump sidebands, which contain a few percent of the total power, to the cooling light. We then ramp the intensity of the D$_1$ light to $ 10~I_{\textrm{s,D}_1}$ and its detuning to $2.9~  \Gamma_{\textrm{D}_2}$ over $4~$ms. This reduces the temperature of the Li cloud to $ 30~\upmu$K and preserves $\sim 75\%$ of the  atom number initially in the MOT.


The initial laser cooling and trapping of erbium broadly follows established techniques~\cite{Aikawa:2012becErbium}. 
The Er MOT operates on the yellow intercombination line at $583~$nm and employs a magnetic field gradient of $2.6~$G$/$cm along the strong axis. 
The cooling light is red-detuned by $44~\Gamma_{583}$, where $\Gamma_{583}=2\pi\times 186~$kHz is the natural linewidth of the 583 nm transition, and we broaden the spectrum of the MOT light by $\sim 3~$MHz. 
The beams each have a waist of $14~$mm, and in the horizontal plane have a central intensity of $220~I_{\textrm{s},583}$, where $I_{\textrm{s},583}=0.13~$mW$/$cm$^2$ is the saturation intensity of the 583 nm transition. 
The vertical MOT beam operates with a lower central intensity of $20~I_{\textrm{s},583}$ since it provides a negligible contribution to the slowing and capture of the atomic beam. We typically load $~7\times10^7$ atoms in $8~$s, after which the MOT beam intensities are equalized, spectral broadening is ceased, and the laser detuning is set to $38~\Gamma_{583}$. 
We then further cool the trapped atoms by reducing the MOT beam central intensity to $0.05~I_{\textrm{s},583}$ over $\sim 45~$ms and then holding for $420~$ms. Simultaneously, the detuning is ramped to $21~\Gamma_{583}$ and the quadrupole gradient to $1.1~$G$/$cm. This reduces the temperature of the cloud to $6.5~\upmu$K, close to the Doppler temperature of $4.6~\upmu$K, and polarizes the atoms in the ground $\ket{J=6,m_J=-6}$ spin state~\cite{Dreon:2017spinPol}.

\subsection{\label{subsec:loadingTransportBeam}Dipole trap loading}
Since the MOTs for trapping erbium and lithium require quadrupole field gradients differing by an order of magnitude, they cannot be operated simultaneously. We therefore sequentially cool each species and load the atoms into an optical dipole trap. This ``transport beam" is formed by a multi-frequency laser at $1050~$nm, with a waist of $47~\upmu$m and a power of up to $100~$W. 
Since Er serves as the coolant in subsequent evaporation, we prioritize its number in the mixture and hence first cool and load Li atoms into the transport beam. 
We find that the presence of the transport beam does not substantially influence the Li MOT or gray molasses cooling, and thus simply overlap it with the quadrupole field center and maintain a constant power of $\sim 60~$W. Upon extinguishing the gray molasses light, $\sim 5\%$ of the Li atoms remain trapped in the transport beam with a temperature of $45~\upmu$K and with an equal distribution across the three lowest hyperfine states $\ket{F=1,m_F=0,\pm 1}$. 
We then optically pump the atoms into the ground $\ket{F=1,m_F=1}$ state to provide stability against spin-changing collisions. This is accomplished by applying a magnetic bias field of $6~$G and a $3~$ms pulse of light which is resonant with the $\ket{F=1}\rightarrow\ket{F'=1}$ D$_1$ transition.
The light is circularly polarized such that $\ket{F=1,m_F=1}$ is a dark state in which the atoms accumulate. To avoid any population of atoms in the $\ket{F=2}$ manifold, we apply a small amount of light resonant with the $\ket{F=2}\rightarrow\ket{F'=3}$ D$_2$ transition.

The Er MOT is then switched on, and we simultaneously spatially broaden the transport beam by a factor of three and reduce its power to $17~$W. This yields a trap depth of $61~\upmu$K for Er and $100~\upmu$K for Li. 
The large detuning of the Er MOT light means that atoms accumulate at a location $\sim 1.6~$cm below the quadrupole field center~\cite{Dreon:2017spinPol}, and do not overlap with the trapped Li cloud. 
After compression of the Er MOT, we ramp up a small vertical bias field over $360~$ms to overlap the Er cloud with the transport beam, and typically load $~13\times 10^6$ atoms into the dipole trap at a temperature of $\sim6~\upmu$K. After loading, we maintain a bias field of $500~$mG to define a spin quantization axis, cease spatial broadening of the transport laser, and increase its power to $60~$W. This yields a trap depth of \kb$\times 640~\upmu$K and \kb$\times 1~$mK, and trap frequencies of $2\pi\times(1200,1200,6)~$Hz and $2\pi\times(7500,7500,38)~$Hz for Er and Li, respectively. 

\subsection{\label{subsec:transport}Optical transport and trapping in the glass cell}

After loading, we transport the atomic mixture to a glass cell via a mechanical translation of the transport beam focus. The duration of the transport is $2~$s, and we observe minimal atom loss or heating associated with motion of the trap. We then switch on a crossed optical dipole trap (cODT) consisting of a single-frequency $1064~$nm beam arranged in a bow-tie configuration.
This beam has a power of $10~$W and a waist of $40~\upmu$m on the first pass and $30~\upmu$m on the second, and it is polarized in-plane to avoid interference between the two passes. 
The trap depths are $390~\upmu$K and $635~\upmu$K, and the trap frequencies are $2\pi\times({669,1146,1260})~$Hz and $2\pi\times(4040, 6920, 7620)~$Hz for Er and Li, respectively. 
Due to the weak axial confinement provided by the transport beam, the spatial overlap of the atomic cloud with the cODT is small. We therefore first load atoms from the transport beam into a counter-propagating, multi-frequency laser with a wavelength of $1035~$nm, power of $90~$W, and waist of $32~\upmu$m focused at the center of the glass cell. This increases the central atomic density by a factor of three. We then ramp down this beam in the presence of the cODT. 
Midway through the rampdown, we apply a $100~$ms stage of Doppler cooling~\cite{Schmidt:2003doppler} provided by a single beam of $\sigma^-$-polarized $583~$nm light, with an intensity of $\sim \textrm{I}_{\textrm{s},583}$ and red-detuned by $1.44~\Gamma_{583}$, propagating parallel to a magnetic bias field of $0.8~$G. 
This enhances the Er atom number loaded into the cODT by a factor of $\sim 1.6$, yielding $1.8\times 10^6$ Er atoms and up to $5\times 10^4$ Li atoms, both at a temperature of $32~\upmu$K. Around $10^6$ Er atoms are confined in the central region of the cODT, with the remainder trapped in the wings. At this stage, the central phase space density of both species is $\sim 10^{-3}$.

\section{\label{sec:evaporation}Evaporative cooling}

The next stage of the experiment is forced evaporative cooling to degeneracy. In Sec.~\ref{subsec:evapErOnly}, we describe the cooling of Er in the absence of Li, while Sec.~\ref{subsec:evapMixture} discusses cooling of the mixture. We find that sympathetic cooling, aided by fast thermalization of the two species, is very efficient. In Sec.~\ref{subsec:scattLength}, we present a measurement of the interspecies scattering length between Er and Li at low magnetic field.

\begin{figure}[h]
\includegraphics[width=\columnwidth]{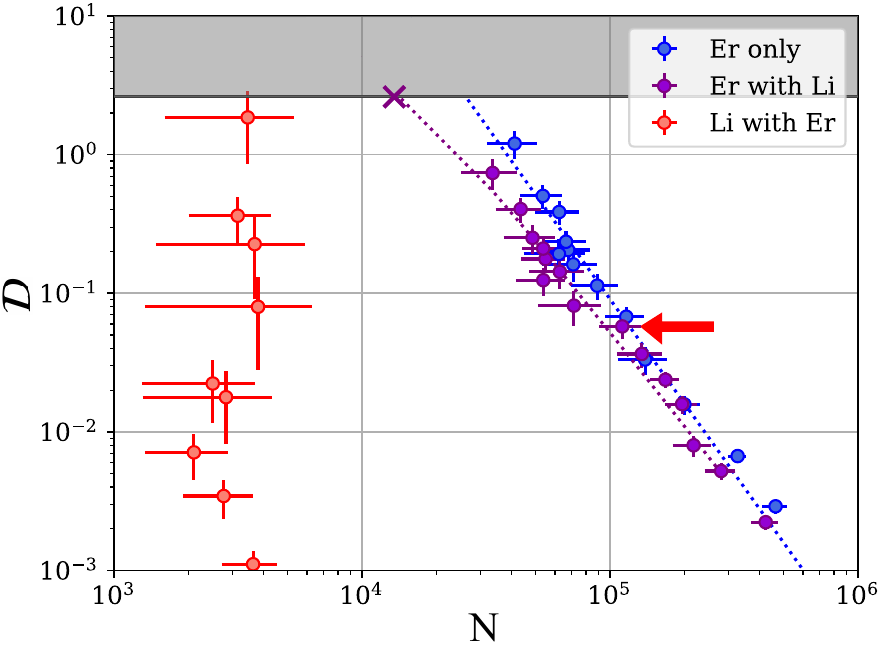}
\caption{Evolution during evaporative cooling of the phase space density at the cloud center, $\mathcal{D}$, with atom number in the central region of the optical trap, $N$. The black line indicates the threshold for condensation. Data correspond to an Er cloud in the absence of Li (blue), Er in the presence of Li (purple), and Li in the presence of Er (red). A pure erbium cloud exhibits an efficiency parameter of $\sim 3$, indicated by a dashed blue line. The addition of Li minimally affects the initial Er evaporation, but causes the efficiency to drop as the number of Er atoms decreases. This trend is well captured by a simple model (see text) shown by a dashed purple line. The observed onset of Er condensation is indicated by a purple cross. The Li cloud exhibits an increase in $\mathcal{D}$ by approximately three orders of magnitude with minimal detectable atom loss. Owing to its smaller atomic mass, Li condenses much earlier than Er (see text), indicated by a red arrow. Each point shows the average of $11$ iterations of the experiment, and error bars are statistical.}
\label{fig:sympatheticcooling}
\end{figure}

\subsection{\label{subsec:evapErOnly}Cooling a pure erbium cloud}
Erbium is evaporatively cooled via an exponential rampdown of the cODT power by a factor of $24$, with a duration of $17.5~$s and a time constant of $3.6~$s.
In Fig.~\ref{fig:sympatheticcooling}, we show the evolution of the phase space density at the cloud center, $\mathcal{D}$, with decreasing atom number in the central region of the optical trap, $N$. After an initial detrimental atom loss, which we attribute to the imperfect spin purity of the cloud after transport, we observe evaporation of Er with an efficiency parameter $-\mathrm{d}(\log\mathcal{D})/\mathrm{d}(\log N)\approx 3$. Evaporation is found to be optimal at a bias field of $1.3~$G, consistent with prior results for this isotope~\cite{Krstajic:2023}.
At the end of this ramp, $\sim 10^5$ atoms remain at a temperature of $630~$nK, with a geometric mean trapping frequency of $2\pi\times 200~$Hz and $\mathcal{D}\sim 0.3$. 
Further decompression of the cODT would result in an isotropic or prolate trap, which is unfavorable for lanthanide condensates~\cite{lahaye2009dipolarphysicsreview}. Furthermore, at this stage the gravitational sag of Er is comparable to its thermal radius, and further evaporation would result in spatial separation from a co-trapped Li cloud.
To compensate for this, we ramp up a sheet beam propagating in the plane of the cODT while linearly ramping the cODT down to a final depth of \kb $\times2.9~\upmu$K over $2.3~$s. 
The sheet beam has axial and transverse waists of $8.9~\upmu$m and $480~\upmu$m respectively and a final depth of \kb$\times2.1~\upmu$K.
Together with the cODT, this yields final trapping frequencies of $2\pi\times(57,98,390)~$Hz. At the end of this ramp, we achieve condensates of $\sim 1.5\times 10^4$ Er atoms, depicted in Fig.~\ref{fig:dualBEC}. 

\subsection{\label{subsec:evapMixture}Cooling of the mixture}
The evaporation sequence in the presence of Li is unchanged, and in Fig.~\ref{fig:sympatheticcooling} we show the phase space evolution for both species. Here, the number of Li atoms loaded into the cODT is $\sim 3\times 10^3$.
While the initial evaporation trajectory of Er is essentially unchanged, the falling number of coolant atoms is accompanied by a corresponding reduction in efficiency. 
This trend is captured by a simple model of sympathetic cooling dynamics developed by the Aspect group~\cite{delannoy2001understanding} and whose prediction is indicated by a purple dashed line. 
The only free parameter of this fit is the typical energy $\approx 6.3\times$\kb $T$ of an Er atom lost from the trap, and we fix the initial atom number and temperature such that the curve is anchored to a data point in the early stages of evaporation.\!~\footnote{In principle, condensation of the Li cloud is accompanied by an initial increase in its heat capacity, which then decreases toward zero as the condensate fraction grows. By modifying the model of~\cite{delannoy2001understanding} to account for this, we find that these two effects essentially compensate one another, and our data remain well-described by assuming a constant heat capacity throughout.}
The Li cloud exhibits extremely efficient evaporation throughout, with $\mathcal{D}$ increasing by approximately three orders of magnitude with minimal associated atom loss.

We observe that condensation of Li occurs significantly earlier than Er in the evaporation sequence, indicated by a red arrow in Fig.~\ref{fig:sympatheticcooling}. 
This is a consequence of the larger polarizability and smaller mass of the Li atoms. 
For a given optical trap power, and in the case of a thermalized mixture with equal atom numbers held in a harmonic trap, the central phase space density of a thermal gas of Li is $(\alpha_\textrm{Li}m_\textrm{Er} / (\alpha_\textrm{Er}m_\textrm{Li}))^{3/2}\approx 240$ times greater than that of Er.
The ratio of polarizabilities in our mixture is fortuitous, since it means that the heavier Er atoms serve as the coolant and remain thermal throughout condensation of Li. On the other hand, if the coolant atoms were to condense first, their associated reduction in heat capacity would inhibit further sympathetic cooling.

After further evaporation, Er also condenses and crosses the critical temperature with $\sim 10^4$ atoms remaining, indicated by a purple cross in Fig.~\ref{fig:sympatheticcooling}.
This yields a final condensed number of $1\;\textrm{--}\;4\times 10^3$ for each species. 

\subsection{\label{subsec:scattLength}Scattering length determination}

The thermalization dynamics of a strongly mass-imbalanced mixture differ in several ways from the mass-balanced case. First, the energy exchanged per collision is reduced by a factor of $(\mEr+\mLi)^2/(4\mEr \mLi)\approx 6.5$~\cite{Mudrich:2002sympCooling}, and second, the gravitational sags of the two species differ by a factor $\alpha_\textrm{Li}m_\textrm{Er}/(\alpha_\textrm{Er}m_\textrm{Li})\approx 39$, eventually leading to a loss of spatial overlap between the clouds. In our experiment the gravitational sag of Er reaches a maximum of $ 4.4~\upmu$m just before the sheet beam is ramped up, comparable to the thermal radius of $3.8~\upmu$m for Er and $3~\upmu$m for Li.
However, we do not observe any benefit from increasing the duration of the evaporation ramp in the presence of Li, indicating that the limiting thermalization process remains Er-Er collisions.
This is explained by two factors. First, the smaller mass of Li means that its thermal speed is enhanced by a factor of $\sqrt{\mEr/\mLi}\approx 4.9$ compared to Er, which increases the relative speed of Li-Er pairs with respect to Er-Er. 
Second, we measure a relatively large interspecies scattering length of $a_{\textrm{ErLi}} = 100 \pm 10~a_0$, resulting in rapid thermalization of the Li cloud with the Er bath.

\begin{figure}[h]
\includegraphics[width=\columnwidth]{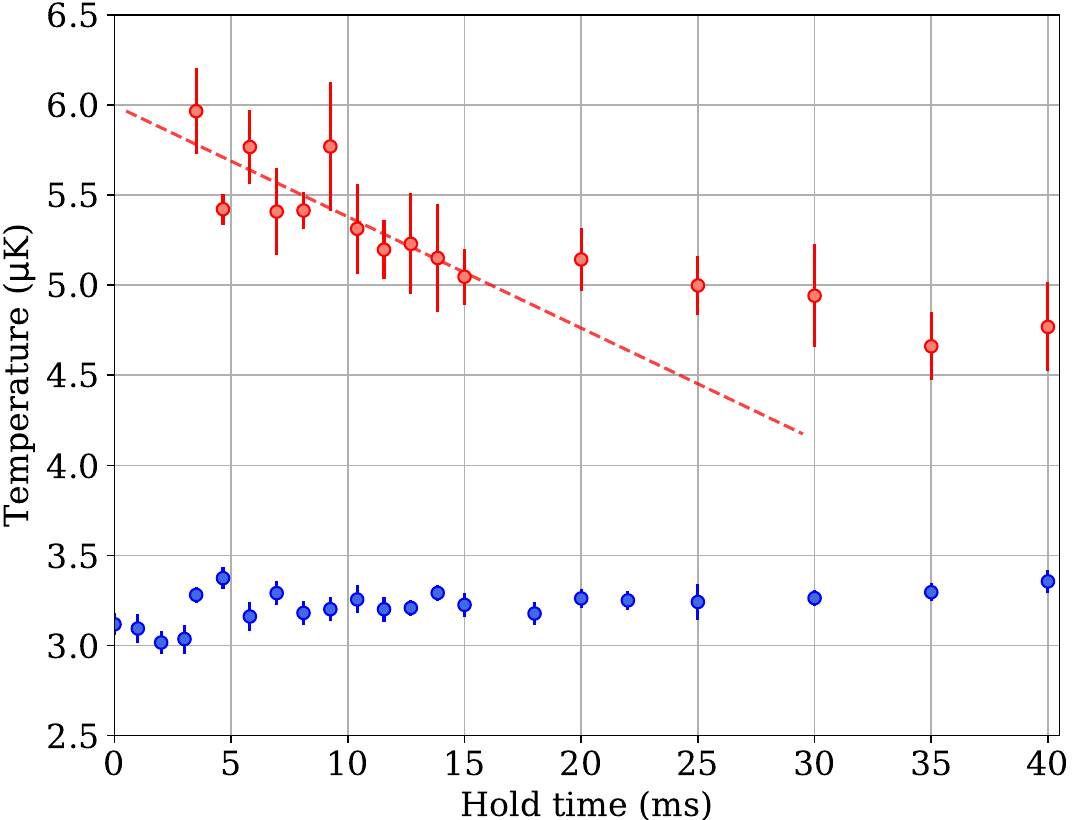}
\caption{Interspecies thermalization of $^{166}$Er and $^7$Li. We use a resonant light pulse with a $5~\upmu$s duration to selectively heat the Li cloud and measure the relaxation of the two species towards thermal equilibrium. By fitting the data to a simple thermalization model (see text), we infer an interspecies $s$-wave scattering length of $a_{\textrm{ErLi}} = 100 \pm 10~a_0$ at $1.3$ G.}
\label{fig:thermalization}
\end{figure}

To extract the interspecies scattering length, we perform a thermalization measurement at the magnetic field of $1.3$ G used for evaporation of the mixture. We prepare a mixture of $2.9\times 10^4$ Er atoms and $1.8\times 10^4$ Li atoms at a temperature of $3.3~\upmu$K, confined in an optical trap with trapping frequencies $2\pi\times (237, 405, 445)~$Hz for Er and $2\pi\times (1430, 2450, 2690)~$Hz for Li. 
We then selectively heat the Li cloud by several microKelvin via a 5~$\upmu$s pulse of light with $\sigma^+$ polarization and resonant with the $\ket{F=2}\rightarrow\ket{F'=3}$ D$_2$ transition. Since the intraspecies scattering length of Li is small~\cite{Hulet:2009icap}, this predominantly heats the Li cloud along one axis and makes an analytic treatment of the resulting temperature dynamics of the two clouds difficult. Furthermore, differential evaporation and background heating rates can lead to offsets between the temperatures at long times~\cite{guttridge2017interspecies}.
We therefore adopt a simple approach, and extract the effective temperature describing the momentum distribution of the Li atoms along the heated axis as a function of time. This is plotted in Fig.~\ref{fig:thermalization}, and shows a decrease toward the Er temperature which remains approximately constant.
We perform a linear fit to the early time evolution, shown by the dashed line, and assume that the energy exchange is mediated by elastic scattering between the Li and Er atoms. From a simple model, the details of which are given in Appendix~\ref{appendix:thermalization}, we extract a scattering length of $a_{\textrm{ErLi}} = 100 \pm 10~a_0$.

\section{\label{sec:Conclusion}Conclusion and outlook}

We have demonstrated the simultaneous production of Bose-Einstein condensates of $^{166}$Er and $^7$Li atoms in their ground spin states. After sequential cooling and loading of each species into an optical trap, we sympathetically cool the Li cloud via forced evaporation of the Er atoms. The mixture exhibits highly efficient cooling, enabled by a large interspecies elastic scattering length. This establishes Er-Li as a promising platform for studying the interplay of large mass imbalance, strong and tunable interactions, and emergent phases in binary mixtures of dipolar and non-dipolar quantum fluids. Loss spectroscopy recently revealed numerous interspecies Feshbach resonances~\cite{schafer2022feshbach, schaffer2023fermionicEr}, and a natural next direction is to characterize suitable candidates for tuning the interspecies scattering length. 

At present, the principal limitation to the number of atoms brought to degeneracy is the weak trapping of the transport laser along the axial direction. This leads to both a substantial loss of phase space density when atoms expand from the MOT into the optical trap and poor spatial matching of the transported cloud with the cODT in which evaporation is performed. Furthermore, the multi-frequency nature of the transport laser precludes application of a sufficiently large bias field during transport to preserve spin polarization. We are therefore in the process of replacing this setup with a moving optical lattice formed from a single-frequency laser~\cite{schmid2006opticallattice}.

\appendix
\section{Description of the apparatus}
\label{appendix:apparatus}
%
We employ a commercial effusion oven for Er (CreaTec DFC-40-10-WK-2B). The aperture is formed from an array of $3$D-printed titanium micronozzles of length $10~$mm and diameter $200~\upmu$m, which limits the angular spread of the emitted atomic beam to $\lesssim 20~$mrad. 
The Er beam then passes through a transverse cooling stage, formed by a retro-reflected beam of $401~$nm light with waists $13.3~$mm and $3.5~$mm along and perpendicular to the atomic beam, respectively, a central intensity of $1.1~I_{\textrm{s},401}$, where $I_{\textrm{s},401}=60.3~$mW$/$cm$^2$ is the saturation intensity of the $401~$nm transition, and red-detuned by $0.5~\Gamma_{401}$, where $\Gamma_{401} = 2\pi \times 29.7$ MHz is the natural linewidth of the $401~$nm transition.
This is followed by a custom chamber which allows $2/3$ of the Er atoms to pass through unimpeded, and contains a Li oven localized to the lower $1/3$ of the chamber and separated from the Er beam by a wall of thickness $200~\upmu$m. 
The MOTs for both species are located inside a titanium octagonal chamber. The atoms are then optically transported a distance of $24~$cm to a glass cell. The light for the transport laser is derived from a multi-frequency laser (IPG YLR-200-1050-LP-WC), and the light for the cODT and sheet beam in the glass cell from a Azurlight ALS-IR-1064-50-A-CP-SF seeded by a Mephisto 500. 
The atoms in the cell are imaged via a microscope objective with NA=$0.28$. We also have an NA=$0.6$ objective for high-resolution imaging.

\section{Extraction of interspecies scattering length}
\label{appendix:thermalization}
As described in the main text, we extract the interspecies scattering length from a thermalization measurement. The temperature of the Li cloud is suddenly increased, and we observe the relaxation of the two species toward a new thermal equilibrium.
Assuming purely $s$-wave scattering, the rate of energy exchange between the two clouds is described by a simple model~\cite{delannoy2001understanding, ivanov2011sympathetic, guttridge2017interspecies} $\dot U_\text{Er} = -\dot U_\text{Li} =  \xi \overline n_{\textrm{ErLi}}  \sigma\overline v   k_\textrm{B}  (T_\textrm{Li}-T_\textrm{Er}),$ where $U_i$ is the internal energy of species $i$,  $\sigma$ is the interspecies scattering cross section, $\xi = 4 m_\textrm{Er} m_\textrm{Li} /(m_\textrm{Er} +m_\textrm{Li})^2$ is the correction factor for unequal masses \cite{Mudrich:2002sympCooling}, $\overline n_{\textrm{ErLi}} = \int n_\textrm{Er} (\boldsymbol{r}) n_\textrm{Li}(\boldsymbol{r}) \textrm{d}^3 \boldsymbol{r}$ is the number of collision partners per unit volume, and $\overline v = \sqrt{8 k_\textrm{B} (T_\textrm{Er}/m_\textrm{Er}+T_\textrm{Li}/m_\textrm{Li})/\pi } $ is the mean relative thermal velocity. The temperature difference $\Delta T = T_\textrm{Li}-T_\textrm{Er}$ then obeys
\begin{equation}
\frac{\text{d}\Delta T}{\text{d}t} = \left(\frac{\dot U_\text{Li}}{3 k_\textrm{B} N_\textrm{Li}} -\frac{\dot U_\text{Er}}{3 k_\textrm{B} N_\textrm{Er}}  \right) = -\frac{\xi \overline n \sigma \overline v }{\alpha} \Delta T,
\label{eq:therm}
\end{equation}
where $\overline n = ({1}/{N_\textrm{Er}}+{1}/{N_\textrm{Li}}) \int n_\textrm{Er}(\boldsymbol{r}) n_\textrm{Li}(\boldsymbol{r}) d^3 \boldsymbol{r}$ is the overlap density, and the average number of collisions needed to thermalize, $\alpha=3$, follows from the heat capacity $3 N k_B$. 

The solution to Eq.~\ref{eq:therm} is in general not exponential due to the dependence of $\overline v$ and $n(\boldsymbol{r})$ on the temperature of each species. Furthermore, any differential heating and cooling of the two species, for example due to evaporation or single-photon scattering, should in principle be included~\cite{guttridge2017interspecies}. 
We therefore take the Er temperature as constant and perform a linear fit to the Li temperature at short times to extract the initial cooling rate, according to the following equation:\begin{equation} \frac{d  T_\text{Li}}{dt}  \approx - \frac{\xi \sigma}{\alpha}   \langle \overline n \rangle \langle \overline v \rangle (\langle T_\text{Li}\rangle - \langle T_{\text{Er}} \rangle)  ,  \end{equation} where the angled brackets indicate quantities averaged over the early times used for the fit. The only free parameter is the cross section $\sigma = 4\pi a^2$, from which we extract the interspecies scattering length $a$. 

We note that Er and Li atoms may also exchange momentum via elastic dipolar scattering; however, following~\cite{ravensbergen2018production} we calculate the corresponding cross section to be $\sim 1.6\times 10^{-20}~$m$^2$ and thus negligible. 
Additionally, the temperature of the atoms is far below the $p$-wave scattering threshold $\sim\sqrt{4/C_6}(\hbar^2/(3\mu))^{3/2} \approx k_\textrm{B}\times 2.3 ~$mK~\cite{ivanov2011sympathetic}, where $C_6$ for the Er-Li system was calculated in~\cite{gonzalez2015erliFeshbach} and $\mu$ is the reduced mass. 

While this model is derived assuming an isotropic momentum distribution of both species, in our experiment the heating of Li is not isotropic. We do not expect this to change the applicability of Eq.~\ref{eq:therm}, since the separability of the Boltzmann distribution implies that each axis of an isotropically heated sample cools independently of the others when in contact with a cold reservoir. However, when evaluating the density distribution of the Li cloud, we explicitly account for the different effective temperature along each axis, and, when evaluating $\overline v$, we replace $T_\textrm{Li}$ with the average of the effective temperatures characterizing the momentum distribution along each axis. 

\begin{acknowledgments}
We thank Sarah Wattellier, Mathis Demouchy, Ruoyi Yin, and Zhenjie Yan for experimental contributions in the early stages of the experiment, and Peter Schauss, Rob Smith, and Christian Gross for useful discussions. 
This work was supported by the NSF through the Center for Ultracold Atoms (PHY-2317134)
and Grant PHY-2207367,
the AFOSR Young Investigator Program (FA9550-22-1-0066),
and the David and Lucile Packard Foundation (2023-76156).
This material is based upon work supported by the
Defense Advanced Research Projects Agency (DARPA) under Agreement No. HR00112490527.
J.K. and J.R. acknowledge support from  the National Science Foundation Graduate Research Fellowship under Grant No. 2141064.

\end{acknowledgments}

\bibliography{refs}

\end{document}